\documentclass[12pt]{article}

\usepackage{graphicx}
\usepackage{epsf}

\usepackage{amsmath,amssymb,graphicx, float} 

\usepackage{pstricks}
\usepackage{color}

\def\beq{\begin{equation}}
\def\eq{\end{equation}}
\def\eeq{\end{equation}}

\def\centeron#1#2{{\setbox0=\hbox{#1}\setbox1=\hbox{#2}\ifdim
\wd1>\wd0\kern.5\wd1\kern-.5\wd0\fi
\copy0\kern-.5\wd0\kern-.5\wd1\copy1\ifdim\wd0>\wd1
\kern.5\wd0\kern-.5\wd1\fi}}
\def\ltap{\;\centeron{\raise.35ex\hbox{$<$}}{\lower.65ex\hbox{$\sim$}}\;}
\def\gtap{\;\centeron{\raise.35ex\hbox{$>$}}{\lower.65ex\hbox{$\sim$}}\;}

\def\lsim{\mathrel{\ltap}}

\def\MET{{\not \!  \! E}_T}
\def\METsub{{\not \!  E}_T}
\def\subMET{{\not \!  E}_T}\def\projMET{{\not \!  \!E}_T^{\rm proj}}

\def\chii0{\chi_i^0}
\def\chij0{\chi_j^0}

\def\foursqr#1#2{{\vcenter{\vbox{
 \hrule height.#2pt
 \hbox{\vrule width.#2pt height#1pt \kern#1pt
 \vrule width.#2pt}
 \hrule height.#2pt
 \hrule height.#2pt
 \hbox{\vrule width.#2pt height#1pt \kern#1pt
 \vrule width.#2pt}
 \hrule height.#2pt
     \hrule height.#2pt
 \hbox{\vrule width.#2pt height#1pt \kern#1pt
 \vrule width.#2pt}
 \hrule height.#2pt
     \hrule height.#2pt
 \hbox{\vrule width.#2pt height#1pt \kern#1pt
 \vrule width.#2pt}
 \hrule height.#2pt}}}}
\def\psqr#1#2{{\vcenter{\vbox{\hrule height.#2pt
 \hbox{\vrule width.#2pt height#1pt \kern#1pt
 \vrule width.#2pt}
 \hrule height.#2pt \hrule height.#2pt
 \hbox{\vrule width.#2pt height#1pt \kern#1pt
 \vrule width.#2pt}
 \hrule height.#2pt}}}}
\def\sqr#1#2{{\vcenter{\vbox{\hrule height.#2pt
 \hbox{\vrule width.#2pt height#1pt \kern#1pt
 \vrule width.#2pt}
 \hrule height.#2pt}}}}

\def\figin{\epsfcheck\figin}\def\figins{\epsfcheck\figins}
\def\epsfcheck{\ifx\epsfbox\UnDeFiNeD
\message{(NO epsf.tex, FIGURES WILL BE IGNORED)}
\gdef\figin##1{\vskip2in}\gdef\figins##1{\hskip.5in}
\else\message{(FIGURES WILL BE INCLUDED)}%
\gdef\figin##1{##1}\gdef\figins##1{##1}\fi}
\def\DefWarn#1{}
\def\figinsert{\goodbreak\midinsert}
\def\ifig#1#2#3{\DefWarn#1\xdef#1{fig.~\the\figno}
\writedef{#1\leftbracket fig.\noexpand~\the\figno}%
\figinsert\figin{\centerline{#3}}\medskip\centerline{\vbox{\baselineskip12pt
\advance\hsize by -1truein\noindent\footnotefont{\bf
Fig.~\the\figno:\ } \it#2}}
\bigskip\endinsert\global\advance\figno by1}

\def\fig#1#2#3#4{\vskip 0.5cm \begingroup \midinsert \centerline{
\psfig{file=#1,width=#2}} \vskip 0.4cm
\global\advance\figno by 1
\centerline{\vbox{\baselineskip=12pt \noindent Figure \the\figno: #3}}
\endinsert \endgroup {\xdef#4{\the\figno}} }

\def\figcrop#1#2#3#4#5#6#7#8{\vskip 0.5cm \begingroup \midinsert \centerline{
\psfig{file=#1,width=#2,bbllx=#3,bblly=#4,bburx=#5,bbury=#6}} \vskip 0.4cm
\global\advance\figno by 1
\centerline{\vbox{\baselineskip=12pt \noindent Figure \the\figno: #7}}
\endinsert \endgroup {\xdef#8{\the\figno}} \vskip .5cm}

\def\figlabel#1{\xdef#1{\the\figno}}
\def\encadremath#1{\vbox{\hrule\hbox{\vrule\kern8pt\vbox{\kern8pt
\hbox{$\displaystyle #1$}\kern8pt}
\kern8pt\vrule}\hrule}}
\def\underarrow#1{\vbox{\ialign{##\crcr$\hfil\displaystyle
 {#1}\hfil$\crcr\noalign{\kern1pt\nointerlineskip}$\longrightarrow$\crcr}}}

\begin{document}

\begin{titlepage}

\begin{center}
\vspace*{-1cm}

\hfill RU-NHETC-2011-16 \\
\hfill UTTG-20-11 \\
\hfill TCC-022-11 \\
\vskip 1.0in
{\Large \bf Backgrounds to Higgs Boson Searches from   } \\
\vspace{.15in}
{\Large \bf $W\gamma^* \to \ell \nu \ell (\ell)$ Asymmetric Internal Conversion }
\vspace{.15in}


\vskip 0.65in
{\large Richard C. Gray}$^1$~
{\large Can Kilic}$^{1,2}$~
{\large  Michael Park}$^1$~
\vskip 0.1in
{\large Sunil Somalwar}$^1$~
{\rm and} ~
{\large Scott Thomas}$^1$

\vskip 0.25in

$^1${\em 
Department of Physics \\
Rutgers University \\
Piscataway, NJ 08854}

\vskip 0.15in
$^2${\em 
Theory Group, Department of Physics and Texas Cosmology Center\\
The University of Texas at Austin \\
Austin, TX 78712}

\vskip 0.75in
\end{center}

\baselineskip=16pt

\begin{abstract}
\noindent
A class of potential backgrounds for Higgs boson searches
in the $h \to WW^{(*)} \to \ell \nu ~\ell^{\prime} \nu$ channel
at both the Tevatron and Large Hadron Collider is presented.
Backgrounds from $W \gamma$ production
with {\it external} conversion of the on-shell photon
in detector material to an asymmetric electron--positron pair, $\gamma \to e(e)$,
with loss of the trailing electron or positron has been treated
adequately in Higgs searches.
Here we consider analogous backgrounds
from $W \gamma^*$ production with
{\it internal} conversion of the off-shell photon in vacuum to an
asymmetric lepton--anti-lepton pair
$\gamma^* \to \ell (\ell)$.
While the former process yields almost entirely electrons
or positrons, the
latter can give electron, positron,
muon, and anti-muon
backgrounds in roughly equal amounts.
We estimate that asymmetric internal conversion backgrounds
are comparable to the Higgs boson signal
in the standard signal region of phase space.
These processes also represent potential backgrounds for
new physics searches in same-sign di-lepton channels.
Some data driven methods to characterize asymmetric internal
conversion backgrounds are suggested.

\end{abstract}

\end{titlepage}

\baselineskip=17pt

\newpage





\section{Introduction}

The search for the Higgs boson
has been the cornerstone of the physics program at modern high
energy colliders.
The Higgs boson of the Standard Model has well defined
production and decay modes that allow for mass dependent
searches in a number of channels.
One of the key discovery modes at hadron colliders is Higgs boson
production by gluon-gluon fusion
with decay through two leptonically decaying
$W$-bosons, $gg \to h \to WW^{(*)} \to \ell \nu ~\ell^{\prime} \nu$,
giving opposite sign di-leptons plus missing energy.
The dominant background in this channel comes from electroweak
pair production of $W$-bosons, $q \bar{q} \to WW\to \ell \nu ~\ell^{\prime} \nu$.
This background is substantially larger than the Higgs boson signal.
However, the two processes have somewhat different kinematic
properties that may be exploited using either cut based or multi-variate
techniques.
Based on the expected kinematic properties of the signal
and dominant di-boson background obtained from simulations,
searches in this channel have been carried out at both the
Tevatron \cite{Aaltonen:2010cm,Abazov:2010ct,Aaltonen:2010yv}
and Large Hadron Collider (LHC)  \cite{Chatrchyan:2011tz,Aad:2011qi}.

In addition to the background from $W$-boson pair
production, there are a number of other important
processes that contribute background to the
opposite sign di-lepton plus missing energy channel.
While smaller than the dominant background, some can
be comparable to the Higgs boson signal.
Among these are a class of backgrounds arising from direct
electroweak production of a $W$-boson in association with some other object
that is mis-reconstructed as a fake lepton.
This includes a $W$-boson produced
along with jets, where a jet fakes a lepton,
$q^{'} \bar{q} \to Wj \to \ell \nu ~\ell^{\prime}$.
Another in this class is production of a $W$-boson and photon,
with the on-shell photon undergoing an asymmetric external
conversion to an
electron--positron pair
in the electromagnetic field of
an atomic nucleus within the detector material,
$q^{'}\bar{q}\to W\gamma \to \ell \nu ~ e(e)$, where the
parentheses indicate the trailing electron or positron.
If the conversion is sufficiently asymmetric in momentum,
the trailing member of the pair is not reconstructed
as an independent object and does not ruin the isolation
criteria of the leading one, and
the converted photon fakes an electron or positron.
These backgrounds are treated in ongoing Higgs
boson searches [1--5].

\begin{figure}[t]
\begin{center}
\includegraphics[scale=0.80]{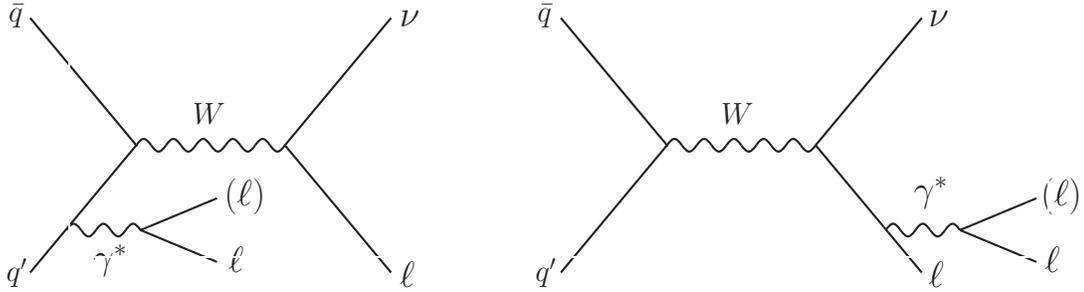}
\vspace{-6.0in}
\caption{Diagrams for production of a leptonically
decaying on/off-shell $W$-boson at a hadron collider
in association with an
initial or final state virtual  off-shell photon radiation
that internally converts in vacuum to a lepton--anti-lepton pair.
Parentheses indicate asymmetric internal conversion
in which the trailing converted lepton is not reconstructed
as an independent isolated object.
Diagrams with an off-shell photon radiated from the
intermediate $W$-boson near the production or decay
vertex are not shown.}
\label{fig:W_conv_fig}
\end{center}
\end{figure}

Here we consider a closely related process within
this class of backgrounds coming from direct production of a $W$-boson
and virtual off-shell photon that undergoes an
internal asymmetric conversion in vacuum
to a lepton--anti-lepton pair,
$q^{'}\bar{q}\to W\gamma^* \to \ell \nu ~ \ell^{\prime}(\ell^{\prime})$,
where $\ell^{\prime} = e, \mu, \tau$.
Initial and final state virtual photon radiation contributions to this process are
shown in Fig. \ref{fig:W_conv_fig}, with
additional contributions coming from $W$-boson virtual
photon radiation near the production or decay vertex.
In a manner similar to the external conversions discussed above,
if the momentum sharing of the conversion pair is sufficiently asymmetric,
the trailing member is not reconstructed as
an independent object and does not ruin the isolation
criteria of the leading one,
and the internal conversion fakes a lepton or anti-lepton.
This process may be referred to as Loss of a Muon or Electron
Following an Asymmetric Internal Conversion (LAME FAIC).

It is instructive to compare and contrast lepton--anti-lepton
pairs arising from external and internal conversion.
In both cases in order for the conversion to give rise to a
single fake object that is reconstructed as a lepton,
the conversion must be sufficiently asymmetric
as described above.
This effective restriction to the asymmetric region of
phase space implies that only a fraction of the
conversions yield fake lepton objects.
Simultaneous reconstruction of a conversion pair
with
both the lepton and anti-lepton identified could recover
most of the remaining
symmetric conversion region of the phase space,
and possibly give a handle on these backgrounds.
Another similarity is that charge conjugation symmetry
of electrodynamics ensures that
conversion photons yield fake leptons of both
charges in roughly equal proportion.
This equality may provide a simple but powerful tool for
characterizing the kinematic properties and distributions
of these backgrounds.
It is already used to constrain the total magnitude of
backgrounds within this class that arise from a $W$-boson in association
with a
mis-reconstructed fake lepton of uncorrelated
charge \cite{Chatrchyan:2011tz,Aad:2011qi}.

External and internal conversions differ in important regards.
The probability for an on-shell photon to convert in material
to a lepton--anti-lepton pair depends strongly on the lepton mass.
Near the forward direction in the high energy asymmetric
limit, the ratio of external conversion
probability for a muon--anti-muon pair
to that for an electron--positron pair scales like
${\cal P}( \gamma \to  \mu \mu)/ {\cal P}( \gamma \to ee) \sim
{\cal O}(m_e / m_{\mu})^2$.
So for all practical purposes external conversions give rise only
to electron--positron pairs.
This is in contrast to internal conversions for which there is
only a moderate logarithmic enhancement of electron--positron over
muon--anti-muon pairs, as described in the next section.
Another key difference is that since external conversion
takes place in material, the reconstructed lepton track in this
case may emerge part-way through the tracking detector.
This feature of missing
hits on the inner part of a reconstructed track
may be utilized as a criterion
for identifying external conversions.
It is however not useful for identifying leptons from internal
conversion since these originate from the collision vertex.

In the next section, we present the theory of asymmetric internal
conversions. Then we study the potential impact of this background on
the Higgs search with a simulation.  Our simulation of the background at the
generator level is done carefully, but the detector simulation that
follows is not particularly sophisticated and is only meant to motivate
detailed studies by the Higgs search teams. We then conclude with a
brief discussion of a possible approach for dealing with the asymmetric
internal conversion backgrounds.
%



\section{Asymmetric Internal Conversion}

The probability for a photon to split to a lepton--anti-lepton pair
by internal conversion
may be calculated in the off-shell photon phase space
using the optical theorem.
The one-loop contribution of a lepton of mass $m_{\ell}$
to the
discontinuity across the branch cut
in the electromagnetic current two-point correlation
function gives the conversion probability distribution
\beq
m_{\ell \ell} ~
{d  {\cal P}( \gamma^* \to \ell \ell) \over d m_{\ell \ell} } =
{2 \over 3} {\alpha \over\pi} \left( 1 - {4 m^2_\ell \over m^2_{\ell \ell} } \right)^{1/2}
\left( 1 +  {2 m^2_\ell \over m^2_{\ell \ell} } \right)
\eq
where $\alpha$ is the fine structure constant, and
$m_{\ell \ell}$ is the lepton--anti-lepton or equivalently off-shell
photon invariant mass.
The internal conversion probability has an infrared soft singularity
in the lepton--anti-lepton invariant mass phase space that is cutoff only
by the lepton mass, $m_{\ell \ell} \geq 2 m_{\ell}$.
The  probability per  logarithmic lepton--anti-lepton
invariant mass in the vicinity of the singularity
is roughly constant.
The total conversion probability integrated between the di-lepton
threshold and an
ultraviolet matching scale $\mu$ is
\beq
{\cal P}( \gamma^* \to \ell \ell) =
\int_{2 m_{\ell}}^{\mu} dm_{\ell \ell} ~
{d  {\cal P}( \gamma^* \to \ell \ell) \over d m_{\ell \ell}} =
{2 \over 3} { \alpha \over \pi} \left[ \ln(\mu / m_{\ell}) - {5 \over 6}
 + {\cal O}(m_{\ell} / \mu)^4 \right]
 \label{splitprob}
\eq
In leading logarithmic approximation, the infrared singular
region of the
lepton--anti-lepton invariant mass phase space gives the dominant
contribution to the process of internal conversion.
The contributions from non-singular regions of
phase space with $m_{\ell \ell} > \mu$ are formally ${\cal O}(1)$
corrections to the logarithm in the brackets in (\ref{splitprob}).
In the background processes of interest here,
the total probability for a high energy photon
to undergo internal conversion to a lepton--anti-lepton pair is
${\cal O}$(1\%).
For example, with $\mu \sim 10$ GeV
the splitting probability
to an electron--positron pair is
${\cal P}( \gamma^* \to ee) \simeq 1.4 \times 10^{-2}$,
to a muon--anti-muon pair is
${\cal P}( \gamma^* \to \mu \mu) \simeq 5.7 \times 10^{-3}$,
and to a tau--anti-tau pair is
${\cal P}( \gamma^* \to \tau \tau) \simeq 1.4 \times 10^{-3}$.

An important kinematic property of
internal conversion is the degree of
asymmetry between the lepton and anti-lepton.
In order to characterize this asymmetry it is useful to define the
momentum fraction carried by the
negatively
charged lepton in the direction of motion of the off-shell photon,
$z = p^{\parallel}_{\ell^-} / (p^{\parallel}_{\ell^+} + p^{\parallel}_{\ell^-})$.
In the high energy co-linear limit
in which the lepton--anti-lepton pair emerges in the direction of the off-shell
photon,
the momentum fraction is related to
the lepton decay angle $\theta$ in the off-shell photon
frame as measured with respect to its direction of motion by
$z = { 1 \over 2} (1 + \beta \cos \theta)$,
where
$\beta = \sqrt{1 - 4 m_{\ell}^2 / m_{\ell \ell}^2 } $
is the lepton velocity in this frame.
The momentum fraction in the co-linear limit lies in the range
$ {1 \over 2} (1 - \beta) \leq z \leq {1 \over 2} (1 + \beta) $.
The probability distribution with respect to this momentum fraction depends on the
polarization of the off-shell photon.
In high energy scattering processes
the probability for emission of an off-shell
photon with longitudinal polarization
is suppressed with respect to transverse polarization by
${\cal O}(m_{\gamma^*} / M)^2$, where $M$
is an ultraviolet mass scale associated with the
hard scattering process.
So for the backgrounds of interest here,
longitudinal polarization may be neglected.

The normalized probability distribution with respect to the
momentum fraction in the co-linear limit for transverse polarization
may be calculated in the two-body lepton--anti-lepton
phase space using the optical theorem as described above,
\beq
{1 \over {\cal P} ( \gamma_T^* \to \ell \ell) }
{d  {\cal P} ( \gamma_T^* \to \ell \ell) \over d z} \equiv f_T(z, \beta) =
{ 2 - \beta^2 +(1 - 2z)^2 \over 2 \beta (1 - \beta^2 /3) }
\eq
Charge conjugation symmetry ensures that this distribution
is invariant under $z \to 1-z$.
Right at threshold, $m_{\ell \ell} = 2 m_{\ell}$,
the conversion probability is symmetric with $z = {1 \over 2}$.
However, well above threshold,
$m_{\ell \ell}^2 \gg m_{\ell}^2 $, the transverse conversion normalized
probability distribution becomes
\beq
f_T(z, \beta) =  {3 \over 4} \left[ 1 + (1 - 2z)^2 \right]
  + {\cal O} (m_{\ell} / m_{\ell \ell})^2
\eq
  with $z$ in the range $0 \leq z \leq 1$.
In this limit  the distribution is maximized for maximally
asymmetric momentum fraction,
$f_T(0,1) = f_T(1,1) = {3 \over 2}$,
and is minimized for symmetric momentum sharing,
$f_T({1 \over 2},1) = {3 \over 4}$.
The total normalized
probability for internal conversion with
momentum fraction $ z, 1-z \leq \epsilon$ in the co-linear limit and
well above threshold is
\beq
\xi(\epsilon) \equiv 2 \int_0^{\epsilon} dz~ f_T(z,1) = 3 \epsilon - 3 \epsilon^2 + 2 \epsilon^3
\eq
This conversion fraction
of course vanishes as $\epsilon \to 0$, but
is not insignificant for moderately small values of
asymmetry.
For example, the fraction of  internal conversions
with asymmetry parameter $\epsilon = 0.15$ is
$\xi(0.15) \simeq 0.39$.
So a sizeable fraction of internal conversions can be
fairly asymmetric.

Another important
kinematic property of internal conversion is the opening
angle between the lepton and anti-lepton.
Near the high energy co-linear limit this angle may be
written in terms of the lepton momentum fraction
and lepton--anti-lepton invariant mass
\beq
\tan \varphi_{\ell \ell} = {m_{\ell \ell} \over 2 |\vec{p}_{\ell \ell}|   }
   {   \sqrt{ \beta^2 - (1 - 2z)^2} \over z(1-z)  }
     + {\cal O}( m_{\ell \ell}^2 /  (|\vec{p}_{\ell \ell}| m_{\ell}))^3
\eq
where $ |\vec{p}_{\ell \ell}| $ is the laboratory frame
total momentum.
The opening angle is small  for
lepton--anti-lepton invariant masses in nearly the
entire range  $ 2m_\ell \leq m_{\ell \ell} \lsim  |\vec{p}_{\ell \ell}| $
for conversions not too far
from the symmetric limit of $z \sim {1 \over 2}$.
The angle is small for all values of momentum fraction,
including near maximal asymmetry
$ z \sim {1 \over 2} ( 1 \pm \beta)$, for invariant masses in the
range
$  2m_\ell \leq m_{\ell \ell} \lsim  \sqrt{ m_\ell |\vec{p}_{\ell \ell}|}$.

%
%
%
%
%
%
%
%
%
%
Before proceeding to an evaluation of the Higgs LAME FAIC background,
we extend Eqn. (\ref{splitprob}). The
differential cross section with respect to any kinematic quantities
$X$ formed from the
four-vectors of the initial and final state particles
including the lepton--anti-lepton pair
is given by
\beq
\int_{2 m_{\ell}}^{\mu} dm_{\ell \ell}~ {d \sigma \over d m_{\ell \ell} dX }
  \big(  {\rm init} \to {\rm final } + \ell \ell \big)
 =
   {\cal P}( \gamma^* \to \ell \ell) \cdot
    {d \sigma \over  dX }
  \big( {\rm init} \to {\rm final }  + \gamma \big)
  \Big|_{p_{\gamma} = p_{\ell \ell}}
  + {\cal O}( \mu/M)^2
\eq
where
${\cal P}( \gamma^* \to \ell \ell)$ is the off-shell
photon to lepton--anti-lepton pair conversion probability
presented in
(\ref{splitprob}), and $M$ is an ultraviolet mass scale associated with the
hard scattering process.



\section{Simulating Asymmetric Internal Conversion}
\label{sec:Simulation}

In order to maximally capture the phase space of the LAME FAIC
asymmetric conversions in a simulation, free parameters in the
simulation package must be chosen with due care.  In addition, the
spatial proximity of the conversion leptons in the tracker can have
large impact on the extent of LAME FAIC background.  The acceptance thus
critically depends on the detector properties, reconstruction algorithms and
kinematic selection criteria.  Therefore, the conclusions of the study that we
describe below are not rigorously quantiative.

In order to simulate 7~TeV LAME FAIC's, we use
Madgraph~(V5)~\cite{Alwall:2011uj}
to separately generate $\ell \nu_{\ell} e^+ e^-$, $\ell \nu_{\ell} \mu^+
\mu^-$ and $\ell \nu_{\ell} \tau^+ \tau^-$ samples.  The Z pole
is removed in order not to doublecount the $WZ$ background. A rapidity cut of $|\eta|<2.5$ was used for all leptons. In
each case, the $p_T$ of the hardest and the second hardest $\ell$ is
required to be above 5~GeV. To capture asymmetric conversions
maximally, $p_T$ of the third lepton is allowed to be as low as
0.1~GeV and the dilepton invariant mass as low as 2$m_{\ell \ell}$.
It was necessary to alter the Madgraph source code
in order to implement these differential thresholds for the daughter
leptons.  The generation cross section for the three processes as
reported by Madgraph are 3878, 1076 and 228~fb for $\ell = e, \mu$
and $\tau$, respectively. For comparison purposes, we also generated
the leptonic decay modes for $gg \rightarrow H$ signal ($m_H =
130~GeV, H \rightarrow WW$) and $qq \rightarrow WW$ Pythia~\cite{Pythia}
samples. The cross section for the Higgs sample was scaled up to the NNLO value \cite{Dittmaier:2011ti}.

These Madgraph and Pythia samples then underwent a generic LHC
detector simulation with the PGS~\cite{pgs} software package. We
altered the PGS source code as follows to implement physics object
isolation in a manner similar to the way it is done at LHC.
An isolation variable is calculated for each photon candidate
centered on an ECAL cell $n$
as a fractional sum of the transverse energy deposited in
ECAL cells surrounding $n$ within a $\eta-\phi$ radius
of $\Delta R = 0.4$
\beq
\gamma_{{\rm iso},n} \equiv
\sum_{ \substack{  {\rm ECAL ~cells } \\  k \neq n} }
{ E_{T,k} |_{\Delta R < 0.4} \over E_{T,\gamma,n} }
\eq
Identified
photons are required to satisfy
$\gamma_{{\rm iso},n} < 0.1$.
This isolation requirement is a good representation of the
full photon identification
requirements used by the LHC experiments.

The sum $p_T$ of all tracks with $p_T$ greater than 0.5 GeV
within an annulus of inner and outer radius
of 0.03 and 0.4 respectively in the $\eta-\phi$ plane
must be less than 0.15 of the $p_T$ of the candidate muon.

\beq
T_{{\rm iso},\mu} \equiv
\sum_{ \substack{ \{ {\rm tracks}~i ~|~ p_{T,i} > 0.5~{\rm GeV},  \\
0.03 \leq \Delta R_{i \mu} \leq 0.4 \}
} } ~
{ p_{T,i} \over p_{T,\mu} }
\eq
$T_{{\rm iso},\mu} < 0.15$

\beq
E_{{\rm iso},e} \equiv
\sum_{ \substack{ \{  {\rm ECAL ~cells } ~k \neq e ~ | \\
3 \times 3 ~{\rm grid}~{\rm surrounding} ~e  \}   }   }
{ E_{T,k}  \over E_{T,e} }
\eq
$E_{{\rm iso},e} < 0.1$.

\beq
p_{T,{\rm iso},e} \equiv
\sum_{ \substack{ \{ {\rm tracks}~i ~| ~
  p_{T,i} > 0.5~{\rm GeV}, \\
  \Delta R_{i e} \leq 0.4 \}
} } ~
 p_{T,i}
\eq
$p_{T,{\rm iso},e} < 5$ GeV

\beq
\rho_e \equiv { E_{T,{\rm ECAL~cell}~e} \over
p_{T,{\rm track}~e}  }
\eq
$ 0.5 \leq \rho_e \leq 1.5$

The total transverse calorimeter energy in a (3 x 3) grid\footnote{We used an ECAL cell size in $\eta$ and $\phi$ of $0.087$, the default value in the PGS parameter set for CMS.} around the
candidate electron (excluding the candidate cell) is defined as
ETISO. PGS then imposes the requirement that
ETISO / $E_t$(candidate) be less than 10\%. The total $P_t$ of
tracks with $P_t >$ 0.5 GeV within a $\Delta R <$ 0.40 cone is defined as
PTISO. In this case, this excludes the leading electron track. It must
satisfy: PTISO $<$ 5 GeV. Finally the ratio of the calorimeter
cell energy to the Pt of the candidate track, E/P, should be within
50\% to 150\%.

Armed with the LAME FAIC event generation followed by a rudimentary detector
simulation, now are in a position to evaluate the Higgs background.
%
%
\section{Di-lepton Plus Missing Energy Background from Asymmetric
Internal Conversion}
\label{sec:background}
We apply a set of selection criteria (``WW Analysis'') that mimic
typical WW selection criteria.  These get the data sample ready for
final kinematic selection and discrimination between the signal and
the background (``Higgs Analysis'').  Table~1 lists these selection
criteria.
%
\begin{table}
\begin{center}
\begin{tabular}{cccc}
\hline \hline
 & &  \\
 & & ~$ee + \mu \mu$ & $e \mu$ \\
&  \underline{$WW$ Analysis}  & & \\ ~
 & & \\
 & $|\eta_{\ell}| ~,~ |\eta_j| $  & \multicolumn{2}{c}{$< 2.5~,~\! < 5$}  \\
 & $ \Delta R_{\ell \ell}$ & \multicolumn{2}{c}{$> 0.4$  } \\
 &  $ p_{T_e} $ & \multicolumn{2}{c}{$> 10$ GeV } \\
 & $ p_{T_\mu} $ & \multicolumn{2}{c}{$> 3$ GeV } \\
  & $ p_{T_j} $ & \multicolumn{2}{c}{$> 30$ GeV } \\
  &   $p_{T_{\ell_1}}, ~p_{T_{\ell_2}} $ & \multicolumn{2}{c}{$> 20~,~ \!10$ GeV}  \\
 & $N_{\ell}~, N_j$  & \multicolumn{2}{c}{Exactly (2,0)}  \\
 &  $Q_{\ell_1} Q_{\ell_2}$ & \multicolumn{2}{c}{$-1$ } \\
 & $\projMET$ & $>40$ GeV & ~~~$>20$ GeV \\
 & $m_{\ell \ell}$ & $>12$ GeV & $-$ \\
 &  $|m_{\ell \ell} - m_Z|$ & $>15$ GeV & $-$ \\
 & & \\
 & \underline{Higgs Analysis}  & & \\
 & & \\
   & $m_{\ell \ell}$ & \multicolumn{2}{c}{$<50$ GeV}  \\
 & $\Delta \phi_{\ell \ell}$ & \multicolumn{2}{c}{$< \pi/2$ }  \\
 & $m_{T,\ell \ell \METsub}$ & \multicolumn{2}{c}{$90-130$ GeV }  \\
  & & \\
 \hline \hline
\end{tabular}
\caption{
Selection criteria and requirements for a representative
opposite sign di-lepton plus missing energy
$W^+W^-$ analysis, along with additional requirements for a
representative 130 GeV Higgs analysis. Jets are defined with a cone algorithm with $\Delta R = 0.5$.
The projected missing transverse energy is defined to be the missing transverse
energy, $\projMET=\MET$ if
$\Delta \phi_{\rm min} > {\pi / 2}$, and to be
$\projMET = \MET \sin(\Delta \phi_{\rm min})$ if $\Delta
\phi_{\rm min} < \pi / 2$, where
$\Delta \phi_{\rm min} =
\min \{  \Delta \phi_{\ell_1 \ \! \! \subMET},\Delta \phi_{\ell_2 \ \! \! \subMET} \}$.
}
\label{analysis}
\end{center}
\end{table}
%

Figs 2, 3, and 4 show distributions of some of the standard kinematics
variables used in Higgs searches.
The $W^+W^-$ analysis selection given in Table~\ref{analysis}
are used for each of the three processes.
In Fig.~5, we show the distribution of transverse mass of the leading
lepton plus missing energy.  This variable is useful for isolating
events with a virtual off-shell photon radiated from the initial state
or the $W$-boson near the production vertex.  This is to be contrasted
with the transverse mass of the both leptons plus missing energy shown
in Fig 4., which is useful for isolating events with a virtual
off-shell photon radiated from the final state lepton or the $W$-boson
near the decay vertex.
It is seen from these figures that at the $W^+W^-$ selection level,
the overall kinematic features of the LAME FAIC background
resemble those of the Higgs rather than those of the WW background. While the
azimuthal angle distribution is very similar for LAME FAIC's and the
Higgs, the mass distributions differ since the former peaks at a lower mass.
\begin{figure}[htbp]
\begin{center}
\includegraphics[scale=0.65]{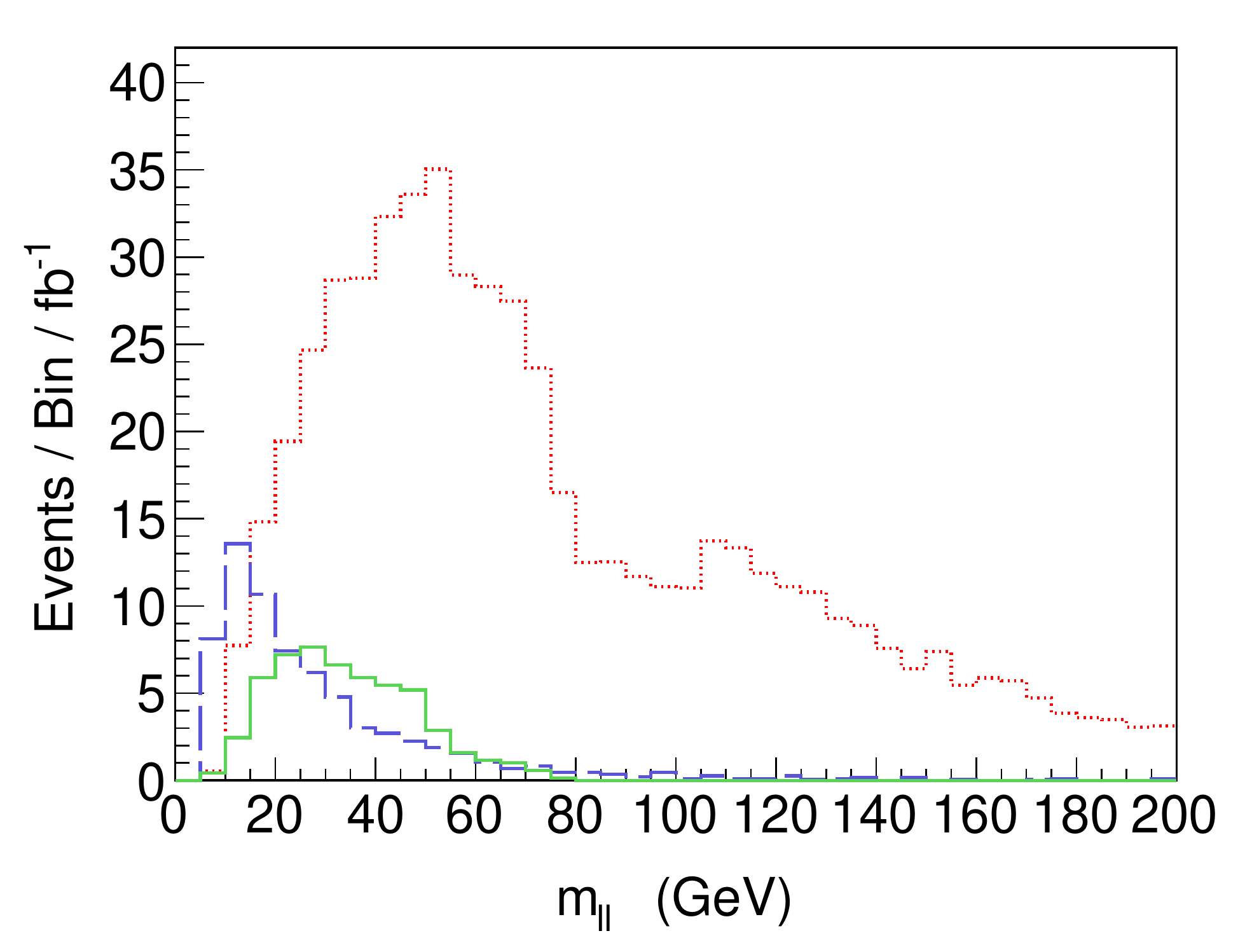}
\caption{Di-lepton invariant mass distributions
arising from $pp$ collisions at 7 TeV
for the $W^+W^-$ analysis selection and requirements
given in Table \ref{analysis} with
PGS acceptance for the
processes
$ q \bar{q}^{'} \to W^+ W^- \to \ell^{+} \nu ~\ell^{'-} \nu$ (red dotted),
$  q \bar{q}^{'} \to W^{\pm} \gamma^* \to \ell^{\pm} \nu ~\ell^{' \mp}(\ell^{' \pm})$
(blue dashed),
and
 $gg \to  h \to W^+ W^- \to \ell^{+} \nu ~\ell^{'-} \nu$
 (green solid)
 with a Higgs boson mass of 130 GeV.
 The leptons $\ell,\ell^{'} = e,\mu$, and $\nu$ refers
to both neutrinos and anti-neutrinos.
The distributions are overlaid, not stacked, with 5 GeV bins. }
\label{fig:mll}
\end{center}
\end{figure}
\begin{figure}[htbp]
\begin{center}
\includegraphics[scale=0.65]{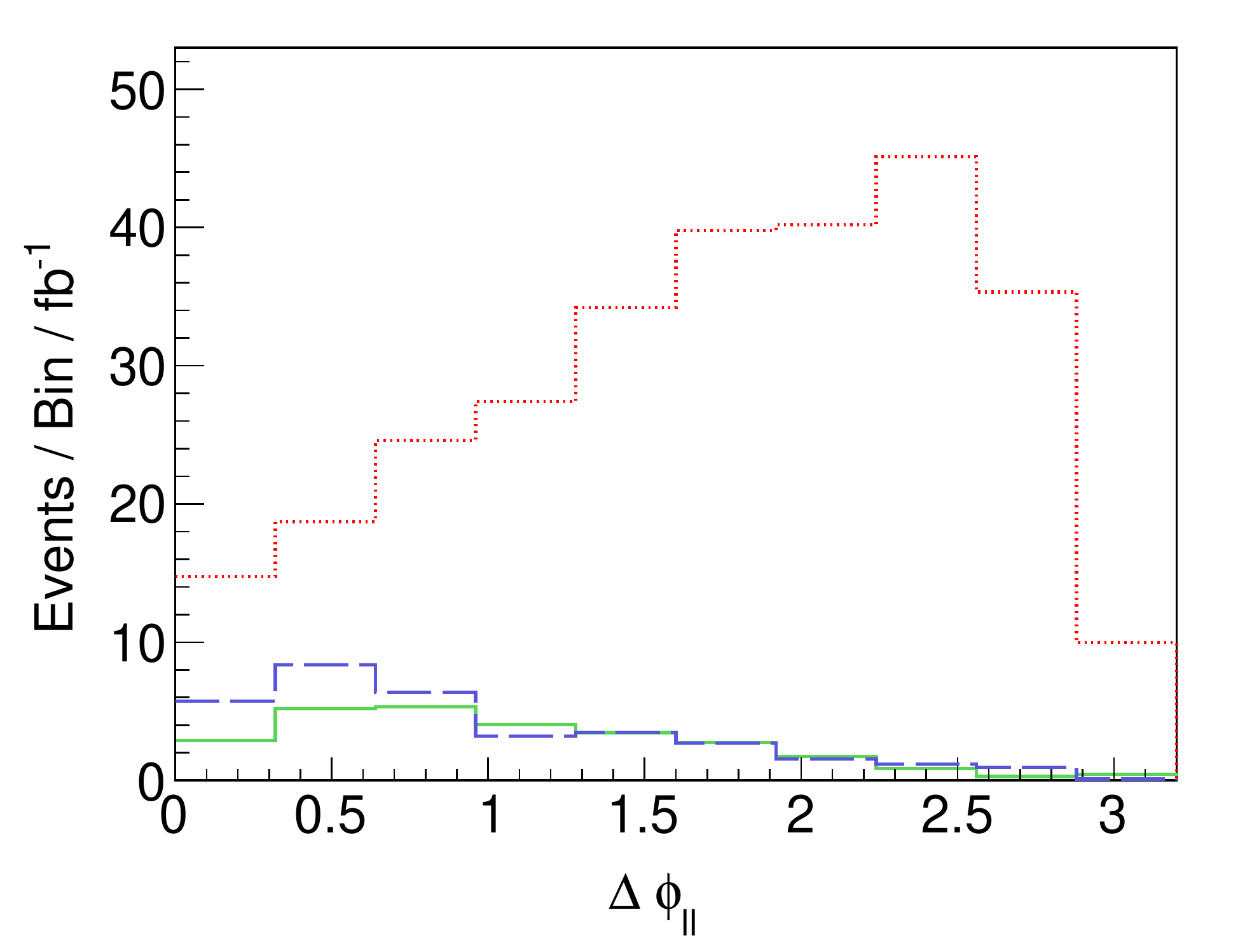}
\caption{Di-lepton azimuthal angle difference distributions
arising from $pp$ collisions at 7 TeV
for the $W^+W^-$ analysis selection and requirements
given in Table \ref{analysis} with
PGS acceptance for the processes
$ q \bar{q}^{'} \to W^+ W^- \to \ell^{+} \nu ~\ell^{'-} \nu$ (red dotted),
$  q \bar{q}^{'} \to W^{\pm} \gamma^* \to \ell^{\pm} \nu ~\ell^{' \mp}(\ell^{' \pm})$
(blue dashed),
and
 $gg \to  h \to W^+ W^- \to \ell^{+} \nu ~\ell^{'-} \nu$
 (green solid)
 with a Higgs boson mass of 130 GeV.
The leptons $\ell,\ell^{'} = e,\mu$, and $\nu$ refers
to both neutrinos and anti-neutrinos.
The distributions are overlaid, not stacked, with bins of size $\pi/10$ radians. }
\label{fig:deltaphi}
\end{center}
\end{figure}
\begin{figure}[htbp]
\begin{center}
\includegraphics[scale=0.65]{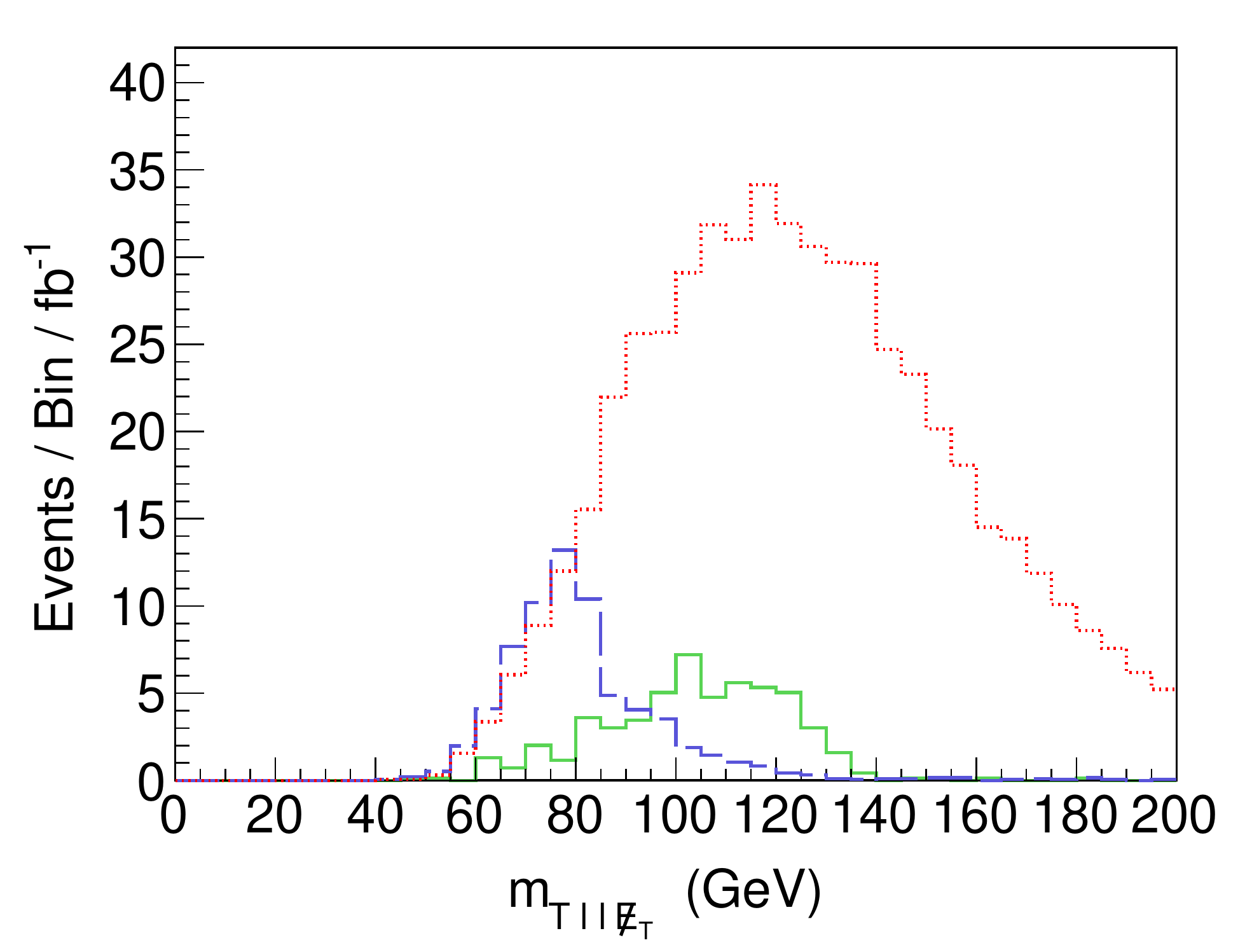}
\caption{Di-lepton plus missing energy
transverse mass distributions
arising from $pp$ collisions at 7 TeV
for the $W^+W^-$ analysis selection and requirements
given in Table \ref{analysis} with
PGS acceptance for the processes
$ q \bar{q}^{'} \to W^+ W^- \to \ell^{+} \nu ~\ell^{'-} \nu$ (red dotted),
$  q \bar{q}^{'} \to W^{\pm} \gamma^* \to \ell^{\pm} \nu ~\ell^{' \mp}(\ell^{' \pm})$
(blue dashed),
and
 $gg \to  h \to W^+ W^- \to \ell^{+} \nu ~\ell^{'-} \nu$
 (green solid)
 with a Higgs boson mass of 130 GeV.
The leptons $\ell,\ell^{'} = e,\mu$, and $\nu$ refers
to both neutrinos and anti-neutrinos.
The distributions are overlaid, not stacked with 5 GeV bins. }
\label{fig:mT}
\end{center}
\end{figure}
\begin{figure}[htbp]
\begin{center}
\includegraphics[scale=0.65]{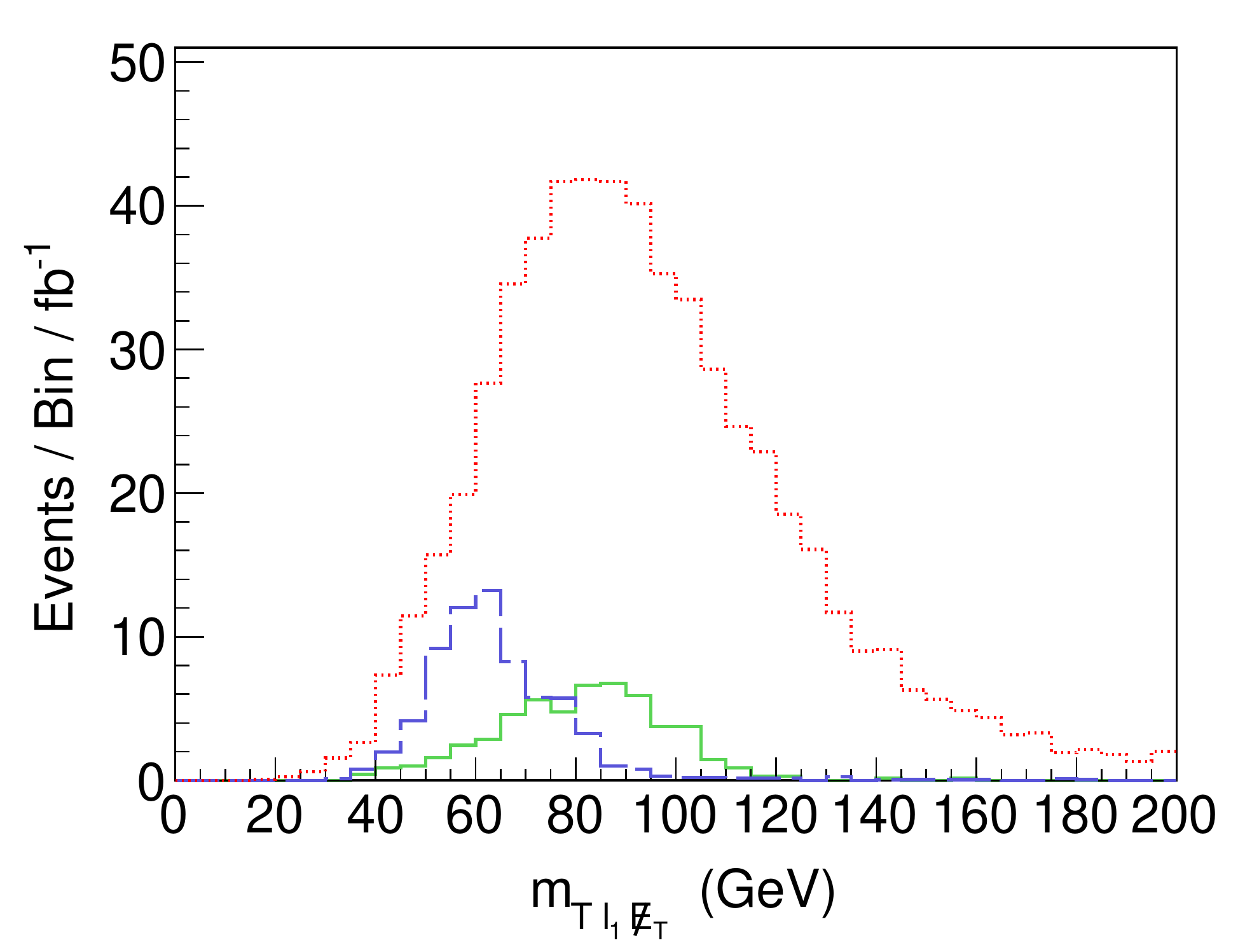}
\caption{Transverse
mass of the leading lepton plus missing energy
distributions arising from $pp$ collisions at 7 TeV
for the $W^+W^-$ analysis selection and requirements
given in Table \ref{analysis} with
PGS acceptance for the processes
$ q \bar{q}^{'} \to W^+ W^- \to \ell^{+} \nu ~\ell^{'-} \nu$ (red dotted),
$  q \bar{q}^{'} \to W^{\pm} \gamma^* \to \ell^{\pm} \nu ~\ell^{' \mp}(\ell^{' \pm})$
(blue dashed),
and
 $gg \to  h \to W^+ W^- \to \ell^{+} \nu ~\ell^{'-} \nu$
 (green solid)
 with a Higgs boson mass of 130 GeV.
The leptons $\ell,\ell^{'} = e,\mu$, and $\nu$ refers
to both neutrinos and anti-neutrinos.
The distributions are overlaid, not stacked with 5 GeV bins. }
\label{fig:mT}
\end{center}
\end{figure}
Dilepton yields for various samples are shown in Table~2.  The first
three rows show the LAME FAIC yields for each (lost) lepton
flavor. The relative flavor decomposition in the table could be
inexact due to the limitations in simulating detector acceptance of an
asymmetric conversion pair.  The fourth row shows the total yield for
the LAME FAIC events.  The last two rows show the same for 130~GeV
Higgs and WW, respectively.  After the WW selection, it is seen that
the LAME FAIC yields are comparable to those of the Higgs even if both are
significantly smaller than the WW yields.
\begin{table}
\begin{center}
\begin{tabular}{lccccccc}
\hline \hline
 & &  \\
%
%
&     \multicolumn{7}{c}{$\sigma \cdot {\rm Br} \cdot {\cal A}_{\rm PGS}$~(fb)  }   \\
 & & \\
   & \multicolumn{3}{c}{$W^{+} W^{-}$ Analysis} &
   & \multicolumn{3}{c}{Higgs Analysis  }  \\
   & $e^{+} e^{-}$  &  $e^{\pm} \mu^{\mp}$  &  $\mu^{+} \mu^{-}$ &
   & $e^{+} e^{-}$  &  $e^{\pm} \mu^{\mp}$  &  $\mu^{+} \mu^{-}$ \\
 & & \\
 $q \bar{q}^{'} \to W^{\pm} \gamma^* \to \ell^{\pm} \nu ~e^{\mp}(e^{\pm})$
     &  7.7 & 47 & 0  &
     &   2.2 & 3.5 & 0         \\

 $q \bar{q}^{'} \to W^{\pm} \gamma^* \to \ell^{\pm} \nu ~\mu^{\mp}(\mu^{\pm})$
     &  0 & 9.9 & 3.4   &
     &  0 & 0.3 & 1.0  \\

 $q \bar{q}^{'}\to W^{\pm} \gamma^* \to \ell^{\pm} \nu ~\tau^{\mp}(\tau^{\pm})$
     &  $<0.1$ & 0.4 & $<0.1$ &
     &  $<0.1$  & $<0.1$ & $<0.1$  \\
      & &  \\

\hline
 & & \\

 $q \bar{q}^{'}\to W^{\pm} \gamma^* \to\ell^{\pm} \nu ~\ell^{' \mp}(\ell^{' \pm})$
     &  7.7 & 57 & 3.4   &
     &  2.2 & 3.8 & 1.0  \\

 $gg \to  h \to W^+ W^- \to \ell^{+} \nu ~\ell^{'-} \nu$
     & 9.4 & 33 & 13   &
     & 7.2 & 16 & 9.4    \\

 $ q \bar{q} \to W^+ W^- \to \ell^{+} \nu ~\ell^{'-} \nu$
     & 90 & 390 & 105 &
     & 21 & 47 & 21 \\
        & &  \\
%
%
%
%
%
%
%
%
\hline \hline
\end{tabular}
\caption{
Cross sections times branching
ratios times PGS acceptance in fb resulting from
$pp$ collisions at 7 TeV
in the $e^+e^-$,
$e^{\pm} \mu^{\mp}$, and
$\mu^+ \mu^-$ channels
for various processes contributing to
the $W^+W^-$ and Higgs
analyses given in Table~\ref{analysis}.
The asymmetric internal conversion processes
do not include  any $Z$ boson contributions.
The penultimate process
includes only gluon fusion production of a 130 GeV Higgs boson,
while the final process does not include
any Higgs boson contributions.
The leptons $\ell,\ell^{'} = e,\mu$, and $\nu$ refers
to both neutrinos and anti-neutrinos.
Parentheses indicate that the lepton is not reconstructed
as an independent isolated object.
}
\end{center}
\label{optable2}
\end{table}
Table~2 also shows dilepton yields after the Higgs selection. As
expected from the accompanying kinematic distributions, the WW
background is reduced drastically. The LAME FAIC's, however, still pose
an ${\cal O}$(20\%) background to the 130~GeV Higgs signal.

\section{Concluding Remarks}
\label{sec:remarks}

We have shown that the internal conversion background is potentially
sizeable and is sufficiently similar in kinematics to the Higgs signal
that it could throw a wrench in the delicate workings of sophisticated
multivariate
analysis techniques employed in the Higgs searches.  Similarities
between internal conversions and Higgs in both yield and kinematic
properties warrants experimental studies of this
background. Furthermore, even sophisticated simulation may prove
inadequate in quantifying detector acceptance of the surviving lepton
since the impact of the lost low-p$_T$ lepton on the tracking and
isolation of the surviving one is difficult to gauge. In that case,
and for the purposes of making searches robust, it behooves the Higgs
hunters to employ data-based techniques for reducing and then
incorporating this background into multivariate schemes.

Since the LAME FAIC background should yield equal number of same-sign
(SS) and opposite-sign (OS) dileptons, the SS dilepton data sample can
be used to constrain the LAME FAIC background in the OS (Higgs)
analysis. A detailed understanding of the ttbar, electroweak and
$W\gamma$ external conversion backgrounds of the SS sample would
constrain the OS LAME FAIC's. A Monte Carlo simulation of the OS LAME
FAIC's that has been validated with the SS data sample could be
integrated in the Higgs multivariate analysis techniques.
%
%
%
%
%
%
%
%
%
%
%
%
%
%
%

To conclude, the opposite-sign dileptons resulting from asymmetric
$W\gamma^*$ internal conversion form a potential background for the
Higgs searches. This background has not been addressed in the recent
results released by the Tevatron or LHC Higgs search teams.  While our
limited study lacks the quantitative rigor in simulating the
intricacies of detector acceptacnce in the presence of a soft
conversion lepton, it is possible that the more detailed simulation
tools currently used by the Higgs search teams are also inadequate in
addressing it, thus necessitating development of data-based
techniques.


\bigskip
\bigskip

{ \Large \bf Acknowledgments}

\smallskip \smallskip

\noindent
We would like to thank Johan Alwall, Emmanuel Contreras-Campana,
Yuri Gershtein,
Amit Lath, Yue Zhao and other colleagues at Rutgers for their insights
and constructive comments.
The research of CK, MP and ST was supported in part by DOE grant
DE-FG02-96ER40959 and NSF Grant PHY-0969020.
The research of RG and SS was supported in part by NSF grant PHY-0969282.


\end{document}